\documentclass[pre,twocolumn,groupedaddress]{revtex4-1}
\usepackage{graphicx}
\usepackage{dcolumn}
\usepackage{bm}
\usepackage{amsmath}
\usepackage{color}
\usepackage{wrapfig}
\usepackage{epstopdf}

\graphicspath{{pics/}}

\newcommand{\om}{\omega}
\newcommand{\bee}{\begin{equation}}
\newcommand{\ene}{\end{equation}}

\newcommand{\bea}{\begin{eqnarray}}
\newcommand{\ena}{\end{eqnarray}}

\renewcommand{\Im}{\mbox{Im}}
\renewcommand{\Re}{\mbox{Re}}

\newcommand{\ve}[1]{{\bf {#1}}}
\newcommand{\te}[1]{{\widehat{#1}}}
\graphicspath{{pics/}}

\let\vec=\mathbf
\begin{document}

\title{Enhancement of inherent Raman scattering in dielectric nanostructures with electric and magnetic Mie resonances }

\author{Kristina Frizyuk$^{1,2}$}\email{k.frizyuk@metalab.ifmo.ru}
\author{Mehedi Hasan$^3$}
\author{Alex Krasnok$^{4}$}\email{akrasnok@utexas.edu}
\author{Andrea Al\'{u}$^{4}$}%
\author{Mihail Petrov$^{1,5}$}

\affiliation{$^1$Department of Nanophotonics and Metamaterials, ITMO University, Saint-Petersburg, Russia\\
$^2$Saint-Petersburg Academic University, Saint-Petersburg, Russia\\
$^3$Division of Physics and Applied Physics, Nanyang Technological
University 637371, Singapore\\
$^4$Department of Electrical and Computer Engineering, The University of Texas at Austin, Texas 78712, USA\\
$^5$Department of Physics and Mathematics, University of Eastern Finland,  80101, Joensuu, Finland}

\begin{abstract}
Resonantly enhanced Raman scattering in dielectric nanostructures has been recently proven to be an efficient tool for developing nanothermometry and experimental determination of their mode-composition. In this paper, we develop a rigorous analytical theory based on the Green's function approach to calculate the Raman emission from crystalline high-index dielectric nanoparticles. As an example, we consider silicon nanoparticles which have a strong Raman response due to active optical phonon modes. We relate enhancement of Raman signal emission to Purcell effect due to the excitation of Mie modes inside the nanoparticles. We also employ the numerical approach to the calculation of inelastic Raman emission in more sophisticated geometries, which do not allow a straightforward analytical form of the Green's function description. The Raman response from a silicon nanodisk has been analyzed within the proposed method, and the contribution of the various Mie modes has been revealed.
\end{abstract}

\maketitle

\section{Introduction}
\label{sec:level1}

Non-plasmonic nanostructures made from high-index dielectrics and semiconductors have recently attracted a great interest, owing to their low-loss optical response and electric and magnetic Mie resonances in the visible spectrum range~\cite{Kuznetsov2016, Jahani2016, Staude2017, Krasnok2015}. Such nanostructures have recently revealed many features, which previously were available only for the plasmonic counterparts. Examples include  single-molecule sensing~\cite{Regmi2016, Yavas2017}, efficient harmonic generation~\cite{Camacho-Morales2016, Timpu2017, Krasnok2017a}, and, among others, photothermal activity~\cite{Zograf2017}. This progress has stimulated the development of all-dielectric photonic devices, including nanoantennas, metasurfaces and optical interconnects~\cite{Staude2017, Kuznetsov2016, Krasnok2015}. However, plasmonic structures have a higher level of local electric field enhancement due to the resonant excitation of localized surface modes, and are actively used for enhancing the Raman signal from external sources, e.g., molecules or nanocrystals~\cite{Ding2016, SHARMA201216, Gwo2016a, Huang2015a}. In stark contrast with metal, non-plasmonic materials often support their own internal Raman response, which -- as proven -- can be exploited as an additional degree of freedom for optical characterization~\cite{Huang2015, Alessandri2016, Dmitriev2016a}.

Semiconductors, which are often considered as the main materials for all-dielectric photonics~\cite{Baranov2017a} have strong inherent Raman response due to their crystalline lattice structure. In particular, crystalline silicon (Si) has a sharp Raman line around $520$~cm$^{-1}$~\cite{Yu2010}, caused by the interaction of light with optical phonons. Recently, it has been experimentally demonstrated ~\cite{Dmitriev2016a} that this Raman response can be resonantly enhanced through the excitation of Mie resonances inherent to non-plasmonic nanostructures. In particular, this has allowed for developing efficient nanothermometry~\cite{Zograf2017} based on the Raman emission from high-index dielectric nanoparticles. Nevertheless, the proper theoretical approach for the description of the Raman emission from resonant Mie nanostructures has not been established yet.

In this paper, we develop a rigorous analytical theory based on the Green's function approach for calculation the Raman emission from crystalline high-index dielectric nanoparticles. The paper is organized as follows: In Section~\ref{SecII} we delineate the general theoretical approach for the description of the Raman emission from localized Raman sources based on the Green's function approach. In Section~\ref{SecIII} we apply the proposed method to calculate the Raman emission intensity from Si nanoparticles and derive the analytical expression for Raman emission intensity at the lowest magnetic dipole resonance. Moreover, we discuss the effect of higher order Mie modes excitation. Here we utilize commercial packages for simulation of the Raman emission, which shows a consistent agreement with the analytical results. In Section~\ref{SecIV} we apply the proposed method for numerical calculation of the Raman emission for the Si nanodisk, whose geometry does not allow the simple analytical Green's function approach. We comment on the influence of different resonant modes of the nanodisk on the intensity of Raman emission.

\section{Raman scattering from dielectric nanostructures}
\label{SecII}
\begin{figure}[!t]
	\includegraphics[width=0.9\linewidth]{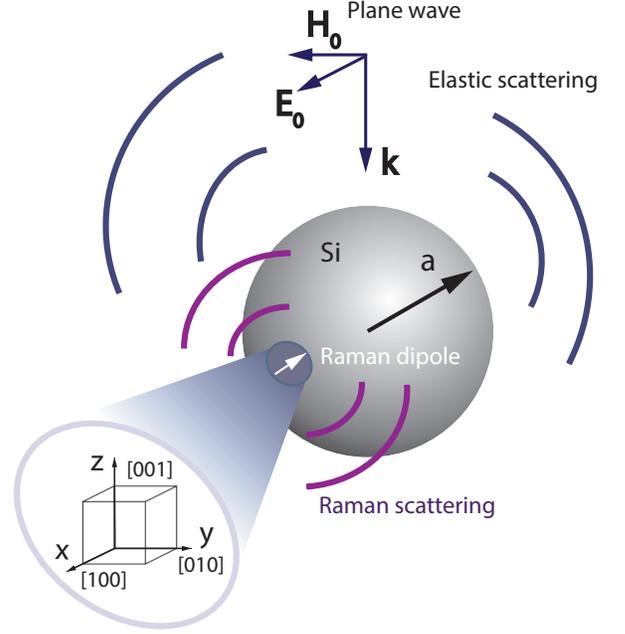}
	\caption{Geometry of the problem. We consider elastic and inelastic scattering of a plane wave by a Si spherical nanoparticle of radius $a=110$~nm in air. The incident field excites the Raman polarization, which generates the Raman signal over the nanoparticle volume. Components of Raman polarization (or Raman dipole moment) are connected with incident field components through the Raman tensor, which is determined by the lattice symmetry properties.}
	\label{Main}
\end{figure}

In this section, we apply the Green's function methodology to develop a model of Raman signal enhancement from a single spherical dielectric nanoparticle of radius $a$ with refractive index $n=\sqrt{\varepsilon}$; $\varepsilon$ is the dielectric permittivity. We assume that the Raman signal is generated by inelastic scattering of a plane wave impinging the nanoparticle along the $z$-axis direction, as shown in Fig.~\ref{Main}. We fix the electric field polarization along the $x$-axis (${\bf E_0}|| x$). Although Raman scattering is a spontaneous quantum process, it allows a classical description, which we employ below. We characterize the lattice vibrations by the phonon coordinate ${\ve Q}(\ve r, t) = {\bf q}(\ve r) \exp(- i\Omega t) $ within the classical description of Raman scattering based on the Raman polarizability tensor~\cite{Fateley1971}. The weak distortions of the crystalline lattice results in the fluctuations of the polarizability tensor $\hat\alpha$ of nanoparticle material allowing the following expansion:
\begin{gather}
 {\alpha_{ij}(\ve r,\omega, \ve Q)=\alpha^0_{ij}(\ve r,\omega )+\sum_k\dfrac{\partial \alpha_{ij}(\ve r,\omega )}{\partial Q_k}\bigg|_{Q_k=0} Q_k(\ve r, t),}
\end{gather}
where $\alpha_{ij}$ are polarizability tensor components and $Q_k$ are the components of phonon coordinates. The first term here stands for the elastic scattering, while the second one is responsible for the generation of Raman polarization. In the spectral representation, the Raman polarization can be described as
\begin{gather}
\label{RamanPol}
\ve P_R(\ve r, \omega, \omega_s)= \te \alpha^R (\ve r, \omega, \omega_s) \ve E_1(\ve r, \omega).
\end{gather}

%
Here $ \alpha^R_{ij} (\ve r, \omega, \omega_s)=\sum_k\dfrac{\partial \alpha^0_{ij}(\ve r, \omega )}{\partial q_k}\bigg|_{q_k=0} q_k(\ve r) $ is the Raman tensor, and $ \omega_s=\omega-\Omega $ denotes that Eq.~(\ref{RamanPol}) stands for the Stokes component of Raman scattering, and $ \ve E_1(\ve r, \omega)$ is the electric field inside the nanoparticle. The Raman tensor is fully defined by the structure of phonon mode spectrum, hence the tensor is defined by the symmetry of crystalline lattice as well as the spectrum of phonon modes. In general, there can be several independent phonon modes in the crystal that have their own phonon coordinates $\ve Q^{\sigma}$, where $\sigma$ labels the phonon mode number. Consequently, the Raman tensor $\hat \alpha ^{R, \sigma}$ should be defined for the each mode $\sigma$.

The total intensity of the Raman signal generated by a particular phonon mode $\sigma$ can be calculated by means of the power flow integration over the entire sphere:
\begin{gather}
\label{inte3}
I_R^{\sigma} = \frac1 {2} \Re\oint [\ve E(\ve r, \omega_s)\times \ve H^*(\ve r, \omega_s)] d\ve S\nonumber \\
=\frac1 {4}\oint [\ve E^*(\ve r, \omega_s)\times \ve H(\ve r, \omega_s)+\ve E(\ve r, \omega_s)\times \ve H^*(\ve r, \omega_s)] d\ve S.
\end{gather}
Here $\ve E(\ve r, \omega_s)$ and $\ve H(\ve r, \omega_s)$ are the electric and magnetic components of the field generated by the Raman polarization $\ve P_R^{\sigma} (\ve r)$ induced by the incident electric field $\ve E_0$. The amplitudes of electric and magnetic fields can be expressed via the Green's function of the system:
\bea \label{fiel}
\vec{\bf E}({\bf r}, \omega_s)&=&\omega_s^2\mu_0\int\limits_V dV' \te {{\bf G}}({\bf r,r'},\omega_s) \ve P_R^{\sigma} (\ve r'),\\
\vec{\bf H}({\bf r},\omega_s)&=&-i \omega_s \int\limits_V dV' \left[\nabla\times \te {{\bf G}}({\bf r,r'},\omega_s)\right] \ve P_R^{\sigma} (\ve r'),
\label{fiel1}
\ena
with $\mu_0$ being the free-space magnetic susceptibility. One can exploit Eqs.~\eqref {fiel} and \eqref {fiel1} in order to rewrite the intensity according to Eq.~(\ref{eq_gf_ir}). With the help of the Green's function properties (see Appendix A), we simplify it even further, obtaining Eq.~(\ref{EmissInt}):
\begin{widetext}
\begin{gather}
\label{eq_gf_ir}
 I_R^{\sigma}
 =\frac{\omega_s^3\mu_0}{4i}\oint\limits_{\partial V,}\iint\limits_{V',V''} d\ve S dV'dV''( \ve{P}_R^{\sigma *}({\bf r'}) \te {{\bf G}}^*({\bf r',r}) \times \nabla \times \te {{\bf G}}({\bf r,r''})\ve{P}_R^{\sigma}({\ve r''}) - \te {{\bf G}}({\bf r,r''})\ve{P}_R^{\sigma}({\bf r''}) \times \nabla \times \vec{P}_R^{\sigma *}({\ve r'}) \te {\ve G}^*({\bf r',r}) ).\\
 I_R^{\sigma} =\frac{\omega_s^3\mu_0 }{2}\left[\ \iint\limits_{V',V''} dV'dV''\ve P_R^{\sigma *}({\bf r'}) \Im\left[\te {{\bf G}}({\bf r',r''})\right]\vec{P}^{\sigma}_R({\bf r''})-k_s^2 \iiint\limits_{V', V'', V} dV' dV'' dV \Im(\varepsilon)\ve{P}_R^{\sigma*}({\bf r'})\te {{\bf G}}^*({\bf r',r})\te {{\bf G}}({\bf r,r''})\ve{P}_R^{\sigma}({\bf r''})\right]. \label{EmissInt}
\end{gather}
\end{widetext}
%


 Eq.~(\ref{EmissInt}) is obtained for a general type of radiative sources distributed over the nanoparticle volume and defined by the polarization vector $\ve{P}_R^{\sigma}(\ve r)$. The Raman scattering is almost a totally incoherent process, and the coherence scale is defined by the phonon propagation length, which is in the order of tens of nanometers for typical materials~\cite{Lim2012, Beams2014}. This allows us to use the ideally incoherent approximation and, thus, we assume that the Raman sources are fully incoherent:
\begin{gather}
\langle \te\alpha^{\sigma'}_R(\vec{r'}, \omega), \te \alpha^{\sigma''}_R(\vec{r''}, \omega)\rangle = \nonumber \\
v\te \alpha^{\sigma'}_R(\vec{r'}, \omega)\te \alpha^{\sigma''}_R(\vec{r''}, \omega) \delta _{\sigma'\sigma''}\delta({\bf r'-r''}),
\label{vol}
\end{gather}
where $\delta _{\sigma'\sigma''}$ is the Kronecker symbol and $v$ is the normalizing volume. Hence, the averaging over the phonon correlations simplifies the expression in Eq.~(\ref{EmissInt}):
\begin{gather}
\langle I_R^{\sigma} \rangle =\frac{\omega_s^3\mu_0 v }{2}\left[\int dV' \ve P_R^{\sigma*}({\bf r'}) \Im\left[\te {{\bf G}}({\bf r',r'})\right]\vec{P}^{\sigma}_R({\bf r'})\right.\nonumber \\
\left.-k_s^2 \iint\limits_{V',V} dV' dV \Im(\varepsilon)\ve{P}_R^{\sigma*}({\bf r'})\te {{\bf G}}^*({\bf r',r})\te {{\bf G}}({\bf r,r'})\ve{P}_R^{\sigma}({\bf r'})\right].
\label{EmissInt2}
\end{gather}
Finally, Eq.~(\ref{EmissInt2}) defines the total emission intensity of the Raman signal from the nanoparticle. The physical meaning of Eq.~(9) is that  the radiated Raman power equals to the total power generated inside the nanoparticle (the first term in Eq.~(\ref{EmissInt2})), and the second term account for the Raman power dissipated in the nanostructure due to the ohmic losses. As we are interested in the Raman emission from all-dielectric materials where the losses are negligibly low, we omit the second term in Eq.~(\ref{EmissInt2}), e.g., the crystalline silicon for wavelengths higher 600~nm range satisfies this requirement.

In order to simplify the expression for Raman intensity even further, we introduce the Purcell factor in terms of the Green's function:
\begin{gather} \label{PurcFac}
F_p^{\sigma}(\omega, \vec r)=\frac{6 \pi c}{\omega} \left[ {\ve n^{\sigma*}_R} \Im\te {{\bf G}}(\omega,{\bf r,r})\ve n^{\sigma}_R \right],
\end{gather}
where $\ve n^{\sigma}$ is the unit vector of Raman polarization $\ve P^{\sigma}_R=P^{\sigma}_R\ve n^{\sigma}_R$. In that way, the intensity of the Raman emission defined by the mode with polarization $\sigma$ can be expressed through the Purcell factor averaged over nanoparticle volume with the weight of the Raman polarization amplitude $\ve P_R^{\sigma}(\ve r)$:
\begin{gather}
\label{EmissFin}
\langle I_R^{\sigma}\rangle=\frac{v \omega_s^4\mu_0}{12 \pi c}\int dV F_p^{\sigma}(\om_s,\vec r) |P^{\sigma}_R( \ve r,\om_s,\om)|^2.
\end{gather}
This expression reflects the spontaneous character of Raman emission. The full incoherence of different phonon modes requires the summation over $\sigma$ in order to obtain the total Raman emission intensity $S_R=\sum_{\sigma}I_R^{\sigma}$. Thus, the intensity of radiation emission is proportional to the square of dipole moment and enhanced by the local density of states, which is defined by the imaginary part of the Green's function. Note that both polarization amplitude $P_R(\ve r,\om_s,\om)$ and Purcell factor can be resonantly enhanced due to the Mie resonances in the all-dielectric nanostructure. However, the resonance of the Raman polarization occurs at the pumping frequency $\om$, as it is defined by the field distribution at the pumping frequency $\ve E_0(\ve r, \om)$ (see Eq.~(\ref{RamanPol})), while the emission of Raman signal occurs at shifted frequency $\om_s$, at which the Purcell factor should be calculated.

\section{Spherical silicon nanoparticles}
\label{SecIII}

We illustrate the results obtained in the previous Section by studying the Raman emission from a Si nanosphere excited by a plane wave. The generated Raman emission is defined by the optical phonon modes of silicon~\cite{Tubino1972} with three orthogonal polarizations $\sigma = x, y, z$. The corresponding modes have different Raman tensors $\te \alpha_{R}^{(x,y,z)}=\alpha \cdot \te R^{(x,y,z)}$, where $\alpha$ is the phonon-polarization independent scalar. Here we fix the orientation of the cubic crystalline lattice according to the coordinate basis (see Fig.~\ref{Main}), which gives us expressions for $R$-tensor~\cite{Yu2010}:
\[
\ \ R^z = \begin{pmatrix} 0 & 1 & 0 \\ 1 & 0 & 0 \\ 0 & 0 & 0 \end{pmatrix}
\ \ R^y = \begin{pmatrix} 0 & 0 & 1 \\ 0 & 0 & 0 \\ 1 & 0 & 0 \end{pmatrix}
\ \ R^x = \begin{pmatrix} 0 & 0 & 0 \\ 0 & 0 & 1 \\ 0 & 1 & 0 \end{pmatrix}.
\]

\begin{figure}[!t]
	\begin{center}
		\includegraphics[width=0.99\linewidth]{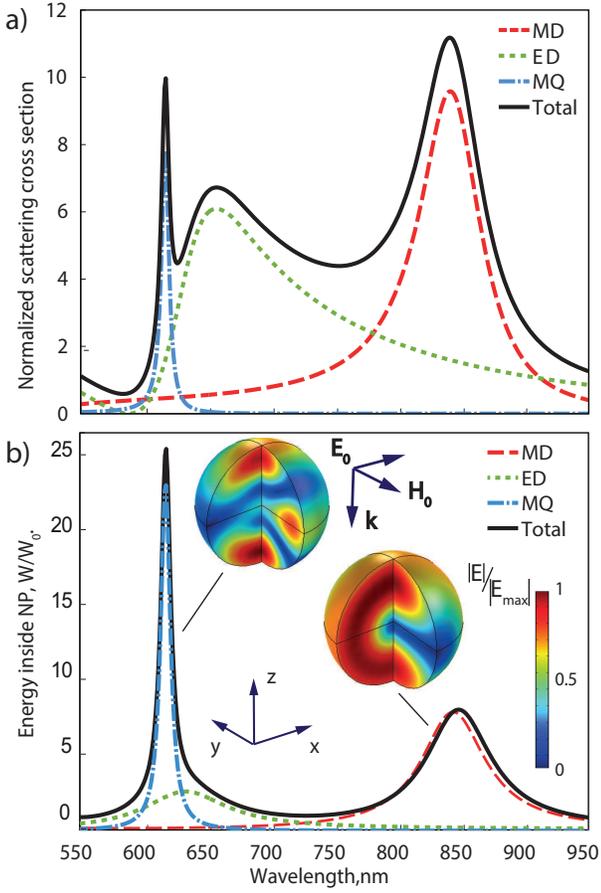}
		\caption{a)~Elastic scattering cross section of the nanoparticle with radius $a=110$~nm. The scattering cross section is normalized over the geometrical one ($\pi a^2$). Mode decomposition with magnetic dipolar (MD), electric dipolar (ED), and magnetic quadrupolar modes (MQ) is also presented. b)~Spectrum of electric energy stored inside the nanoparticle. It is normalized to electromagnetic energy stored in the same volume in a free space $W_0=1/2|E_0|^2$. MD and MQ resonances are plainly visible, however ED does not contribute to the total energy. Insets: the electric field amplitude distribution profiles inside the nanosphere at the MD and MQ resonances.}
		\label{ElasticSphere}
	\end{center}
\end{figure}

In order to study the Raman emission from the Si spherical nanoparticle, we start with the identification of the mode structure by considering elastic scattering of a plane wave. The scattering spectrum obtained within the Mie theory is shown in Fig.~\ref{ElasticSphere} (a). One can clearly see the peaks corresponding to the excitation of the magnetic dipole (MD), electrical dipole (ED), and magnetic quadrupole (MQ) modes. The total scattering cross section is shown along with the contribution of each mode.

The intensity of Raman emission is proportional to the intensity of the electric field inside the sphere. Thus, it is reasonable to plot the electric energy stored inside the sphere, which is defined by the volume integration of the energy density $W=1/{2}\left( |E|^2{d (\omega\varepsilon)}/{d \omega}\right)$. The calculated spectrum is shown in Fig.~\ref{ElasticSphere} (b). One can see that the contribution of the ED mode into the total energy stored inside the nanosphere is low comparing to MD and MQ modes. Thus, one should not expect significant enhancement of the Raman scattering at the ED resonance. The Raman polarization $\ve P_R(\ve r,\om_s,\om)$ can be computed through Eq.~(\ref{RamanPol}), based on the field distribution inside the nanosphere in accordance with Mie theory:
 \begin{gather}
{\bf \vec{E}}_1 = \sum_{n=1}^{\infty} E_n\left( c_n(\om) {\bf \vec{M}}_{o1n}(k_2) - i d_n(\om) {\bf \vec{N}}_{e1n}(k_2)\right) \label{MieIn},
\end{gather}
where ${\bf \vec{N}}_{e1n}$ and ${\bf \vec{M}}_{o1n}$ are vector spherical harmonics, $ E_n=i^nE_0 \frac{2n+1}{n(n+1)}$, $k_2$ is the wavevector taken inside the sphere, the coefficients $c_n$ and $d_n$ are frequency-dependent (see Appendix B).

The Purcell enhancement factor \eqref{PurcFac} can be found through the Green's function of the dielectric sphere~\cite{Li1994}, given by the expressions:
\begin{gather}
	\te {{\bf G}}^{(22)}({\om, \ve r, \ve r'})= \frac {i k_2}{4 \pi } \sum_{n=1}^{\infty} \sum_{m=0}^n (2-\delta_0) \frac {2n+1}{n(n+1)} \frac {(n-m)!}{(n+m)!} \ \times \nonumber \\
 	\times \Bigl( {\bf \vec M}_{^e_o mn}(k_2)\otimes \left[{\bf \vec{M'}}_{^e_o mn}^{(1)} (k_2)+c_n^{(2)}(\om){\bf \vec{M'}}_{^e_o mn}(k_2) \right] + \nonumber \\
	 + {\bf \vec N}_{^e_o mn}(k_2)\otimes \left[ {\bf \vec{N'}}_{^e_o mn}^{(1)} (k_2)+d_n^{(2)}(\om){\bf \vec{N'}}_{^e_o mn}(k_2) \right] \Bigr),
\\ { r < r'}, \nonumber
\end{gather}
\begin{gather}
\te {{\bf G}}^{(22)}({\om,\ve r, \ve r'})= \frac{\vec{e_r}\otimes\vec{e_r}}{k^2}\delta(\vec r-\vec r')+ \nonumber \\
+ \frac {i k_2}{4 \pi } \sum_{n=1}^{\infty} \sum_{m=0}^n (2-\delta_0) \frac {2n+1}{n(n+1)} \frac {(n-m)!}{(n+m)!} \times\nonumber \\
\times \Bigl( \left[{\bf \vec{M}}_{^e_o mn}^{(1)} (k_2)+c_n^{(2)}(\om){\bf \vec{M}}_{^e_o mn}(k_2) \right]\otimes{\bf \vec{M'}}_{^e_o mn}(k_2) + \nonumber\\
+ \left[{\bf \vec{N}}_{^e_o mn}^{(1)} (k_2)+d_n^{(2)}(\om){\bf \vec{N}}_{^e_o mn}(k_2) \right]\otimes{\bf \vec{N'}}_{^e_o mn}(k_2) \Bigr), \label{gfg}
\\ { r \geq r'}. \nonumber
\end{gather}
Here ${\bf \vec{M}}_{^e_o mn}(k_2)$ and ${\bf \vec{N}}_{^e_o mn}(k_2)$ are vector spherical harmonics (see Appendix B) with superscript $(1)$ obtained by replacing spherical Bessel functions $j_n(\rho)$ by spherical Hankel function of the first kind $h_n^{(1)}(\rho)$. We adopt the simplified notation that $ {\bf \vec{N}}_{^e_o mn}(k)\otimes{\bf \vec{N'}}_{^e_o mn}(k)= {\bf \vec{N}}_{omn}(k)\otimes{\bf \vec{N'}}_{omn}(k)+{\bf \vec{N}}_{emn}(k)\otimes{\bf \vec{N'}}_{emn}(k)$ and similarly for other terms; $\delta_0=1$ when $m=0$ and $\delta_0=0$ when $m\neq 0$. 

\begin{figure}[!t]
\begin{center}
\includegraphics[width=0.99\linewidth]{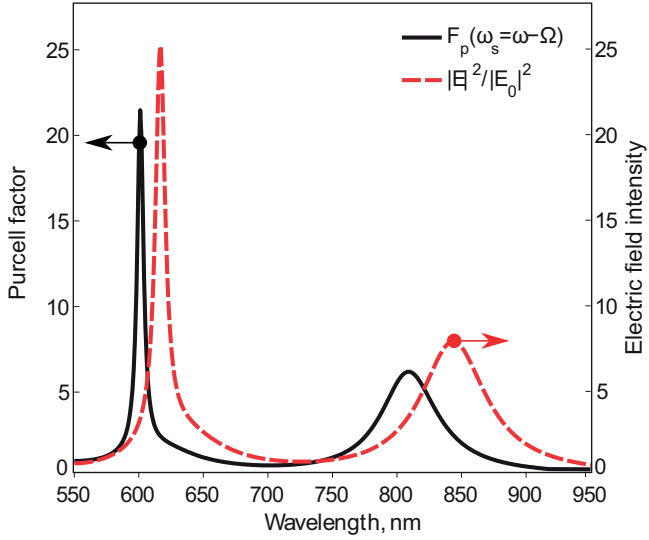}
\caption{Purcell factor (black curve) and electric field intensity (red curve) for the nanoparticle with radius $a=110$~nm. The Purcell factor is averaged over the nanoparticle volume and dipole polarizations. Both factors are multiplied in resulting Raman signal.}
\label{fig:Purcell}
\end{center}
\end{figure}

For comparative analysis of both factors of Raman scattering enhancement, we plot the electric field intensity $|E|^2/|E_0|^2$ spectrum, which is proportional to the electric energy, and the Purcell factor averaged over all polarizations, in Fig.~\ref{fig:Purcell}. The Purcell factor was calculated for the shifted Stokes frequency $\omega_S=\omega-\Omega$. Both quantities demonstrate resonant behavior in the vicinity of Mie resonances.

\begin{figure}[!t]
\begin{center}
\includegraphics[width=0.99\linewidth]{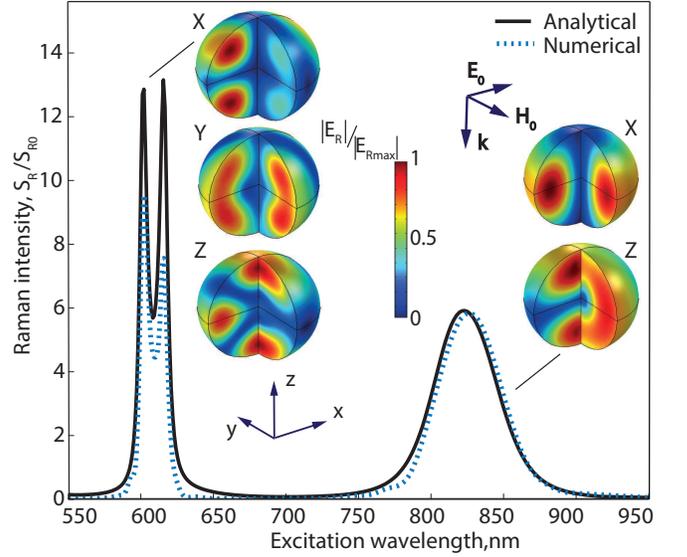}
\caption{Comparison of analytical and numerical results of the incoherent Raman signal intensity spectrum normalized to Raman intensity from the same volume of bulk silicon, given by $S_{R0}={(v \omega_s^4\mu_0 n V |\alpha E_0|^2)}/{(12 \pi c)}$. Inserts: the Raman electric field distribution inside the sphere for X- (up) and Z- (down) phonon modes on the MD resonance, and X- Y- Z- for MQ resonance.}
\label{fig:RamanSphere}
\end{center}
\end{figure}

Next, we perform explicit calculations for the Raman enhancement in the vicinity of the MD resonance taking into account only one term in the expansion of the exciting field and Green's function:
\begin{gather}
	\langle S_R\rangle= \frac{9 v \omega_s^3 k_2 \mu_0}{20}|E_0|^2|c_1(\om)|^2 \times \nonumber \\
	 \left[ \Re \left( c_1^{(2)}(\om_s)\right)\right]\int_0^a r^2 {j_1^4(k_2 	r)} dr .
	 	\label{Emiss_an}
\end{gather}
Both exciting electric field and dyadic Green's function depend on the resonant Mie coefficients ${c_1^{(2)}(\om), d_1^{(2)}(\om), c_1(\om), d_1(\om)}$, which is reflected in the Eq.~(\ref{Emiss_an}). Thus, for a given excitation frequency $\om$ the Raman signal will be resonantly enhanced either when $|c_1(\om)|$ or $c_1^{(2)}(\om_s)$ reaches their maximal values. As shown in Fig.~\ref{fig:RamanSphere} two significant peaks, one of which is shifted by optical phonon frequency, should appear for each resonance. One peak appears when the Raman frequency $\om_s=\om-\Omega$ is resonant, and depicts the enhancement in $c_1^{(2)}(\om_s)$ due to the Purcell factor. The second peak in Fig.~\ref{fig:RamanSphere}, in the vicinity of $600$ nm, occurs when the exciting field is enhanced and $c_1(\om)$ becomes resonant.

The intensity of Raman emission from Si nanoparticle with radius $a=110$~nm for different excitation wavelength given by Eq.~\eqref{EmissFin} is presented in Fig.~\ref{fig:RamanSphere}. Normalizing volume $v$ is defined by the intrinsic phonon correlation length of the material. This picture is in accordance with the spectrum of electric energy at the excitation wavelength shown in Fig.~\ref{ElasticSphere}(a). The double-resonant character is not observed for MD mode as the phonon energy is much smaller than the peak width due to the relatively low Q-factor of the MD mode. However, the splitting for MQ mode, which is high-Q enough, is observed.

In order to support these results, we have developed an approach based on numerical simulation and implemented it using Comsol Multiphysics package. We model the incoherent Raman emission by dividing a sphere into subdomains and defining in each of them the Raman polarization in accordance with the excited field distribution. We consider every subdomain as an elementary Raman dipole and independently calculate the Raman signal from each domain. Afterwards, the intensity from each domain is summed over the whole sphere. The resulting intensity depends on subdomain volume, which is equal to normalizing volume $v$. The results of the numerical modeling are shown in Fig.~\ref{fig:RamanSphere} by the dashed-blue-line. We see that the numerical simulations are in a good agreement with analytical ones. The difference in MQ amplitude is defined by the finite size of the subdomains in the numerical method, with non-uniform Raman polarization $\ve P_R^{\sigma}(\ve r)$ over the domain and finite coherence length, while in analytical computation we have point dipoles distribution, which are fully incoherent. Moreover, there are additional limitations such as the mesh domain size in numerical simulations and the finite size of the element volume of integration of the analytical formula.

Based on these numerical calculations, the distribution of Raman signal inside the nanosphere generated by one subdomain, which is a unit source of Raman emission. The insets in Fig.~\ref{fig:RamanSphere} show the electric field distribution for different excitation wavelength corresponding to MD and MQ resonances for different phonon polarizations. While a plane wave couples to eigenmodes of the sphere with $m=1$, even or odd for different resonances, (see Eq.~\ref{MieIn}), the Raman signal can be coupled to spherical harmonics of any arbitrary azimuthal number. One can also compare the distribution of Raman electric field at the frequency $\omega_s$ with the electric field at the excitation frequency $\om$. Due to the symmetry of the MD polarization and the symmetry of Raman tensor, the field distribution of Raman signal generated by $z$-polarized phonon is rotated with respect to the exciting field.

 This approach allows us to calculate the Raman emission from more complicated nanostructures. In particular, resonant Si nanodisks are very often used as building blocks for all-dielectric photonics devices, including oligomers and metasurfaces as they can be easily fabricated via planar technology.

\section{Raman scattering by silicon nanodisks}
\label{SecIV}

\begin{figure}[!t]
\includegraphics[width=0.99\linewidth]{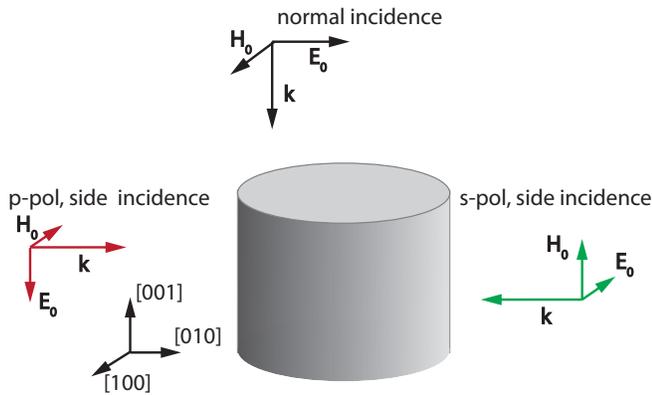}
\caption{Geometry of the problem. We consider elastic and inelastic scattering of a plane wave by a Si nanodisk of radius $a=110$~nm and height $h=190$~nm in vacuum. Three cases of incident field are shown: normal incidence and side incidence with p- and s-polarization.}
\label{MainCylinder}
\end{figure}
\begin{figure}[!t]
\begin{center}
\includegraphics[width=0.99\linewidth]{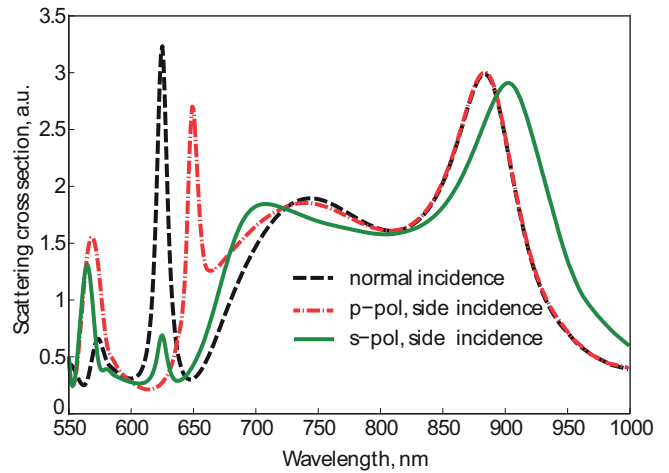}
\caption{Elastic scattering spectrum of the nanodisk with radius $a=110$~nm and height $h=190$~nm for three cases of incident wave: normal incidence, and side incidence for two polarizations. Left peaks (about $900$~nm ) have magnetic dipolar resonance character and peaks between 600~nm and 650~nm are magnetic quadrupolar, by analogy with sphere.}
\label{ElasticCylinder}
\end{center}
\end{figure}

We apply the developed numerical method to compute the intensity of Raman signal generated form a single Si nanodisk with radius $a=110$~nm and height $h=190$~nm, which parameters are close to parameters of the sphere examined in previous section. In case of a spherical nanoparticle the multipole harmonics with different azimuthal number $m$ are degenerate due to spherical symmetry of the problem. The nanodisk has only axial symmetry, thus, the degeneracy is partly lifted. These modes can be excited by a plane wave incident from different directions: normal incidence, and side incidence of $p-$ and $s-$ polarized plane waves, as shown in Fig.~\ref{MainCylinder}. The scattering cross section is depicted in Fig.~\ref{ElasticCylinder} for all three cases. One can see that both MD and MQ modes are spectrally splitted, which is more pronounced for MQ resonance due to higher Q-factor. We leave the discussion of higher order resonances (electric quadrupole, octupole, and etc.) out of the scope of this paper, and they will be elaborated in a subsequent paper.

\begin{figure}[!t]
\begin{center}
\includegraphics[width=0.99\linewidth]{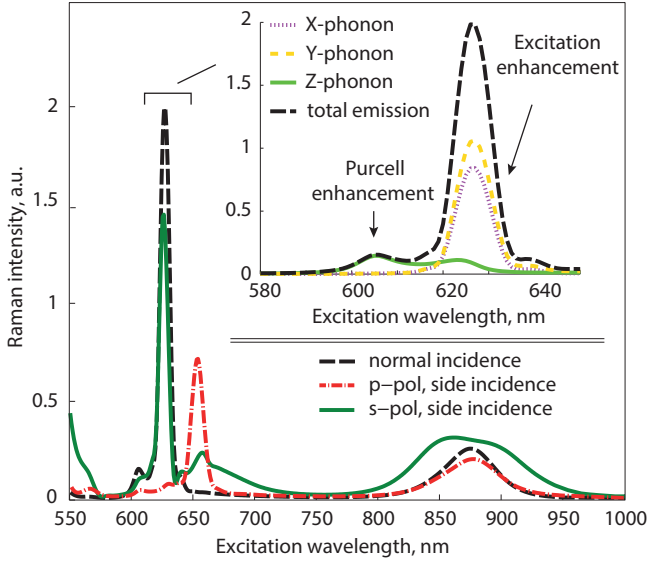}
\caption{Raman scattering by nanodisk with radius $r=110$~nm and height $h=190$~nm spectrum for three cases of incident wave: normal incidence, and side incidence for two polarizations. Insert: detailed phonon-mode decomposition for normal incidence. The Z-phonon has maximal contribution at $\sim 625$~nm when the Purcell effect for left quadrupolar has it's maximal influence. The X- and Y-phonon modes are enhanced when the elastic field inside the sphere is maximal.}\label{RamanCyl}
\end{center}
\end{figure}

The resonant modes observed in the elastic scattering influence the Raman scattering. Although some of these modes may not be seen in the elastic scattering spectrum, they can contribute into overall Raman emission due to breaking of the symmetry by the Raman polarization tensor. Moreover, the resonant peaks can be additionally doubled due to the Purcell factor enhancement similar to the case of nanosphere. In Fig.~\ref{RamanCyl} the Raman intensity is shown as a function of the excitation wavelength. It is obvious that the intensity of Raman signal generation depends on the incidence conditions. However, we also observe a multi-peak structure of the spectrum, which stems from the resonantly enhanced emission by the Purcell effect at different phonon polarizations. In order to analyze it in more details, we circumstantially show the contribution of every phonon mode in the resulting spectrum for normal incidence only (see inset in Fig.~\ref{RamanCyl}). We observe that the total Raman emission curve, shown by the dashed line in the inset, consists of three contributions from different phonon polarization. The largest contribution comes from X- and Y-phonons, with a single peak structure because of the enhancement of the pumping field. There is no Purcell contribution due to specific structure of X- and Y-phonon tensor. For Z-phonon, on the contrary to X- and Y-phonons, we see the active resonant peak at the longer wavelength, due to the Purcell enhancement of the emitted Raman signal, similarly to the splitting shown in Fig.~\ref{fig:RamanSphere}. Moreover, the contribution of second nanodisk MQ mode is faintly noticeable on higher wavelengths, yet weakly enhanced. For a lateral incidence, the mode splitting is also observed for every phonon mode, and more careful analysis of Purcell enhanced response is possible.


\section*{Conclusions}
In conclusion, we have derived a rigorous analytical theory based on the Green's function approach to calculate the Raman emission from crystalline high-index dielectric nanoparticles. For nanoparticles of simple geometry with well-known dyadic Green function, analytical calculations can be performed and significantly simplified in the case of non-absorbing material. It has been demonstrated for the Raman scattering by the resonant Si spherical nanoparticle, as an example. The strongest enhancement has been observed for magnetic resonances, because of the higher field confinement inside the particle. We have also employed the numerical approach to the calculation of inelastic Raman emission in more sophisticated geometries. This approach can be applied for a particle of an arbitrary shape. The Raman response from Si nanodisks has been analyzed within the proposed numerical method and contribution of the various Mie modes has been revealed. We have shown that spectral dependence of the Raman signal intensity from incident light wavelength can reveal modal structure of the particle more clearly than the elastic scattering. The obtained results give the basis for future studies of the resonantly enhanced Raman scattering in high-index all-dielectric nanostructures and its possible applications.

\acknowledgements

We are thankful Andrey Bogdanov, and Alexander Poddubny for very helpful discussions. This work was supported by Russian Foundation for Basic Research proj. \# 16-32-60167. M.P. acknowledges support from Academy of Finland (Grant 310753). K.F. acknowledges support from FASIE.

\appendix
\subsection*{APPENDIX A. Point dipole radiation}

In order to obtain formula \eqref{EmissInt} we use the dyadic Green function formalism~\cite{Novotny2012}:
\begin{gather}
		\nabla_{r''} \times \nabla_{r''} \times \te {{\bf G}}({\bf r'',r})-
		k^2\varepsilon \te {{\bf G}}({\bf r'',r})= \te {{\bf I}}\delta({\bf r'' - r}),			
\end{gather}
where $\te {{\bf I}}$ is unity matrix, multiply by $\vec{\bf P}({\bf r})$ from the right:
\begin{gather}
	 \nabla_{r''} \times \nabla_{r''} \times \te {{\bf G}}({\bf r'',r})\vec{\bf P}({\bf r})	-k^2\varepsilon \te {{\bf G}}({\bf r'',r})\vec{\bf P}	({\bf r})	= \nonumber \\
	 \te {{\bf I}}\delta({\bf r'' - r})\vec{\bf P}({\bf r}),			
\end{gather}
and then multiply by $\vec{\bf P}^*({\bf r'}) \te {{\bf G}}^*({\bf r',r''})$ from the left:
\begin{gather}
\vec{\bf P}^*({\bf r'}) \te {{\bf G}}^*({\bf r',r''})\cdot \nabla_{r''} \times \nabla_{r''}\times \te {{\bf G}}	({\bf r'',r})\vec{\bf P}({\bf r})- \nonumber \\
\label{yde3}
- k^2\varepsilon \vec{\bf P}^*({\bf r'}) \te {{\bf G}}^*({\bf r',r''})\te {{\bf G}}({\bf r'',r})\vec{\bf P}		({\bf r})=\\
=\vec{\bf P}^*({\bf r'}) \te {{\bf G}}^*({\bf r',r''})\te{{\bf I}}\delta({\bf r''- r})\vec{\bf P}({\bf r})	 \nonumber
\end{gather}
Reciprocally for conjugated:
\begin{gather}
\nabla_{r''} \times \nabla_{r''} \times \te {{\bf G^\ast}}({\bf r',r''})-k^2\varepsilon ^* \te {{\bf G^*}}({\bf r',r''})=\te {{\bf I}}\delta({\bf r' - r''})				
\end{gather}
\begin{gather}
\nabla_{r''} \times \nabla_{r''} \times\vec{\bf P}^*({\bf r'}) \te {{\bf G}}^*({\bf r',r''})-k^2\varepsilon^* \vec{\bf P}^*({\bf r'})\te {{\bf G}}^*({\bf r',r''}) = \nonumber \\
=\vec{\bf P}^*({\bf r'})\te {{\bf I}}\delta({\bf r' - r''})\
\end{gather}
\begin{gather}
\nabla_{r''} \times \nabla_{r''} \times\vec{\bf P}^*({\bf r'}) \te {{\bf G}}^*({\bf r',r''})\cdot\te {{\bf G}}({\bf r'',r})\vec{\bf P}({\bf r})-\nonumber \\
\label{yyde3}
 -k^2\varepsilon^*\vec{\bf P}^*({\bf r'})\te {{\bf G}}^*({\bf r',r''})\te {{\bf G}}({\bf r'',r})\vec{\bf P}({\bf r})= \\
\vec{\bf P}^*({\bf r'})\te {{\bf I}}\delta({\bf r' - r''})\te {{\bf G}}({\bf r'',r})\vec{\bf P}({\bf r}) \nonumber
\end{gather}

Subtracting \eqref{yde3} from \eqref{yyde3} and when integrating over $V''$ with vector-dyadic relation transforming volume integration into surface integration we get
\begin{gather}
\vec{\bf P}^*({\bf r'}) \text{Im}\te {{\bf G}}({\bf r',r''})\vec{\bf P}({\bf r''})= \nonumber \\
 = \label{eq_gf} \frac1{2i}\oint\limits_{\partial V} d\ve S( \vec{\bf P}^*({\bf r'}) \te {{\bf G}}^*({\bf r',r}) \times \nabla \times \te{{\bf G}}({\bf r,r''})\vec{\bf P}({\bf r''}) - \\
 - \te {{\bf G}}({\bf r,r''})\vec{\bf P}({\bf r''}) \times \nabla \times \vec{\bf P}^*({\bf r'}) \te {{\bf G}}^*({\bf r',r}) )+ \nonumber \\
 +k^2\int\limits_V dV \text{Im}(\varepsilon)\vec{\bf P}^*({\bf r'})\te {{\bf G}}^*({\bf r',r})\te {{\bf G}}({\bf r,r''})\vec{\bf P}({\bf r''}), \nonumber
\end{gather}
and then substitute \eqref{eq_gf} into \eqref{eq_gf_ir}.

\subsection*{APPENDIX B. Mie scattering amplitudes}

Vector spherical harmonics presented in Mie theory and dyadic Green`s function decomposition:
\begin{gather}
{\bf \vec{M}}_{emn}(k) = \frac {-m}{\sin(\theta)}\sin(m\phi)P_n^m(\cos(\theta))z_n(\rho){\bf \vec e_{\theta}}- \nonumber \\
-\cos(m\phi) \frac{dP_n^m(\cos(\theta))}{d\theta}z_n(\rho){\bf \vec e_{\phi}}
\end{gather}
\begin{gather}
{\bf \vec{M}}_{omn}(k) = \frac {m}{\sin(\theta)}\cos(m\phi)P_n^m(\cos(\theta))z_n(\rho){\bf\vec e_{\theta}}- \nonumber\\
-\sin(m\phi) \frac{dP_n^m(\cos(\theta))}{d\theta}z_n(\rho){\bf\vec e_{\phi}}
\end{gather}
\begin{gather}
{\bf \vec{N}}_{emn}(k) = \frac {z_n(\rho)}{\rho}\cos(m\phi)n(n+1)P_n^m(\cos(\theta)){\bf\vec e_{r}}+ \nonumber\\
+\cos(m\phi) \frac{dP_n^m(\cos(\theta))}{d\theta}\frac1{\rho}\frac{d}{d\rho}[\rho z_n(\rho)]{\bf\vec e_{\theta}}-\\-
m\sin(m\phi) \frac{P_n^m(\cos(\theta))}{\sin(\theta)}\frac1{\rho}\frac{d}{d\rho}[\rho z_n(\rho)]{\bf\vec e_{\phi}} \nonumber
\end{gather}
\begin{gather}
{\bf \vec{N}}_{omn}(k) = \frac {z_n(\rho)}{\rho}\sin(m\phi)n(n+1)P_n^m(\cos(\theta)){\bf\vec e_{r}}+ \nonumber \\+
\sin(m\phi) \frac{dP_n^m(\cos(\theta))}{d\theta}\frac1{\rho}\frac{d}{d\rho}[\rho z_n(\rho)]{\bf\vec e_{\theta}}+\\+
m\cos(m\phi) \frac{P_n^m(\cos(\theta))}{\sin(\theta)}\frac1{\rho}\frac{d}{d\rho}[\rho z_n(\rho)]{\bf\vec e_{\phi}},\nonumber
\end{gather}
where $n=0,1,2, \dots, \ \ \ m=-n,\dots n$, $e$ and $o$ mean two independent solutions for even and odd functions of azimuthal angle. $P_n^m$ are associated Legendre polynomials, $ k^2=\frac{\omega^2}{c^2}\varepsilon, \rho = k a = \frac {2 \pi a }{\lambda}\sqrt{\varepsilon}$, $a$ is nanoparticle radius. $\ve e_{\rho}, \ve e_{\theta},\ve e_{\phi}$ - are unit vectors of spherical basis.

From Mie theory we know fields inside the spherical nanoparticle
 \begin{gather}
{\bf \vec{E}}_1 = \sum_{n=1}^{\infty} E_n\left( c_n(\om) {\bf \vec{M}}_{o1n}(k_2) - i d_n(\om) {\bf \vec{N}}_{e1n}(k_2)\right) \label{decompose}
\end{gather}
\begin{gather}
{\bf \vec{H}}_1 = - \frac{k_2}{\omega \mu_0}\sum_{n=1}^{\infty} E_n\left( d_n(\om) {\bf \vec{M}}_{e1n}(k_2) + i c_n(\om) {\bf \vec{N}}_{o1n}(k_2)\right),
\end{gather}
where ${\bf \vec{N}}_{^e_o mn}$ and ${\bf \vec{M}}_{^e_o mn}$ are vector spherical harmonics, subscript `$2$' of $k$ represents that wavevector is taken inside the sphere, $E_n=i^nE_0 \frac{2n+1}{n(n+1)}$,
\begin{gather}
 c_n(\om) = \frac {\left[ \rho_1 h_n(\rho_1)\right]'j_n(\rho_1) - \left[ \rho_1 j_n(\rho_1)\right]'h_n(\rho_1)}{\left[ \rho_1 h_n(\rho_1)\right]'j_n(\rho_2) - \left[ \rho_2 j_n(\rho_2)\right]'h_n(\rho_1)} \\
 d_n(\om) = \frac {\sqrt{\varepsilon}\left[ \rho_1 h_n(\rho_1)\right]'j_n(\rho_1) - \sqrt{\varepsilon} \left[ \rho_1 j_n(\rho_1)\right]'h_n(\rho_1)}{\varepsilon\left[ \rho_1 h_n(\rho_1)\right]'j_n(\rho_2) - \left[ \rho_2 j_n(\rho_2)\right]'h_n(\rho_1)},
\end{gather}
while coefficients presented in dyadic Green function \eqref{gfg} have different numerator:
\begin{gather}
 c_n^{(2)}(\om) = \frac {\left[ \rho_1 h_n(\rho_1)\right]'h_n(\rho_2) - \left[ \rho_2 h_n(\rho_2)\right]'h_n(\rho_1)}{\left[ \rho_2 j_n(\rho_2)\right]'h_n(\rho_1) - \left[ \rho_1 h_n(\rho_1)\right]'j_n(\rho_2)} \nonumber \\
 d_n^{(2)}(\om) = \frac {n^2\left[ \rho_1 h_n(\rho_1)\right]'h_n(\rho_2) - \left[ \rho_2 h_n(\rho_2)\right]'h_n(\rho_1)}{\left[ \rho_2 j_n(\rho_2)\right]'h_n(\rho_1) - \left[ \rho_1 h_n(\rho_1)\right]'j_n(\rho_2)n^2} \nonumber
\end{gather}
$h_n$ and $j_n$ - are spherical Hankel and Bessel functions,
 \begin{gather}
 \rho_1 = k_1 a = \frac {2 \pi a}{\lambda}, \ \ \
 \rho_2 = k_2 a = \frac {2 \pi a \sqrt{\varepsilon}}{\lambda}.
 \label{tr}
\end{gather}
Scattered fields are:
 \begin{gather}
{\bf \vec{E}}_s = \sum_{n=1}^{\infty} E_n\left( -b_n(\om) {\bf \vec{M}}^{(1)}_{o1n}(k_1) + i a_n(\om) {\bf \vec{N}}^{(1)}_{e1n}(k_1)\right),\\
{\bf \vec{H}}_s = \frac{k_1}{\omega \mu_0}\sum_{n=1}^{\infty} E_n\left( a_n(\om) {\bf \vec{M}}^{(1)}_{e1n}(k_1) + i b_n(\om) {\bf \vec{N}}^{(1)}_{o1n}(k_1)\right),
\end{gather}
where
\begin{gather}
 a_n(\om) = \frac {\varepsilon\left[ \rho_1 j_n(\rho_1)\right]'j_n(\rho_2) - \left[ \rho_2 j_n(\rho_2)\right]'j_n(\rho_1)}{\varepsilon\left[ \rho_1 h_n(\rho_1)\right]'j_n(\rho_2) - \left[ \rho_2 j_n(\rho_2)\right]'h_n(\rho_1)},\\
 b_n(\om)= \frac {\left[ \rho_1 j_n(\rho_1)\right]'j_n(\rho_2) - \left[ \rho_2 j_n(\rho_2)\right]'j_n(\rho_1)}{\left[ \rho_1 h_n(\rho_1)\right]'j_n(\rho_2) - \left[ \rho_2 j_n(\rho_2)\right]'h_n(\rho_1)}.
\end{gather}

\bibliography{AllDRaman_bib}

\end{document}